\documentclass[twocolumn,superscriptaddress,aps,amsmath,amssymb,floatfix,showpacs]{revtex4}

\usepackage{graphicx}
\usepackage{bm}
\usepackage{amsmath}
\usepackage{amsfonts}
\usepackage{amssymb}%
\begin{document}
\title{Charge density wave and superconducting dome in TiSe$_2$ 
from electron-phonon interaction.}
\author{Matteo Calandra}
\author{Francesco Mauri}
\affiliation{CNRS and Institut de Min\'eralogie et de Physique des Milieux
condens\'es, case 115, 4 place Jussieu, 75252, Paris cedex 05, France}
\date{\today}
\begin{abstract}
At low temperature TiSe$_2$ undergoes a charge density wave instability.
Superconductivity is stabilized either by pressure or by Cu intercalation.
We show that the pressure phase diagram of TiSe$_2$ is well described
by first-principles calculations. At pressures smaller than
$4$ GPa charge density wave ordering occurs, in agreement with experiments.
At larger pressures the disappearing of the
charge density wave is due to a stiffening of the short-range force-constants and 
not to the variation of nesting with pressure.
Finally we show that the behavior of T$_c$ as a function of pressure 
is entirely determined by the electron-phonon interaction without need of
invoking excitonic mechanisms. Our work demonstrates that phase-diagrams with competing
orders and a superconducting dome are also obtained in the framework of the 
electron-phonon interaction. 
\end{abstract}
\maketitle
The interplay between  long range charge or magnetic order and
superconductivity is one of the highly debated subject in condensed matter-physics.
Doping of an antiferromagnetic Mott insulator results in high T$_c$
superconductivity in cuprates\cite{MullerHTSC}. 
In iron pnictides \cite{Kamihara2008} magnetism plays
an important role as the parent compound is often an
antiferromagnetic metal with a large magnetic susceptibility \cite{MazinPRL2008}
and spin fluctuations are believed to survive in the superconducting state
\cite{MazinNat2008}.
In transition metal dichalcogenides such as
2H-NbSe$_2$ or 2H-TaSe$_2$, multiband superconductivity \cite{Yokoya2001} and 
charge density wave (CDW) \cite{Revolinski1965} coexist. 

Titanium diselenide (1T-TiSe$_2$)  
belongs to this class of materials, as at $T_{CDW}\approx 200K$
it undergoes a CDW instability characterized by a $2\times 2\times 2$
real space superstructure \cite{MorosanNat2006}. TiSe$_2$ is not
superconducting at low temperature, but CDW is suppressed and
superconductivity stabilized either by Cu intercalation (max. T$_c=4.5$ K)
\cite{MorosanNat2006} or  pressure (max. T$_c=1.8$ K) \cite{KusmartsevaPRL2009}.  
The resulting phase diagram looks  
similar to that of cuprates with the difference that
the spin-density wave order has been replaced by a CDW one\cite{MorosanNat2006}.

The mechanism at the origin of the CDW and superconducting phases in TiSe$_2$ 
is currently unknown. The similarity with the phase diagram of other correlated
materials and the semimetallic \cite{Li07}
nature of TiSe$_2$ led to the idea
\cite{Li07,Cercellier07} 
that the transition from the high-T semimetallic state to the low-T
semimetallic ordered stated is of the Overhauser type\cite{Overhauser}, namely
a Bose-Einstein condensation of excitons driven by correlation effects. 
Indeed, as proposed in refs. \cite{Kohn1967,Halperin}, in a semimetal 
a charge unbalance is poorly
screened due to the lack of carriers and the formation of excitons  
is thus favoured. A Jahn-Teller mechanism has also been
proposed\cite{Hughes, Motizuki} to account for CDW in TiSe$_2$
and in dichalcogenides in general \cite{Whangbo}. This claim
is supported by thermal diffuse
scattering data \cite{Holt}
showing the occurrence of a soft-phonon mode at zone boundary.
However the origin of the softening is unclear as
no first-principles calculations of the TiSe$_2$ phonon dispersion and
electron-phonon coupling are available. It is then 
unknown if density functional theory (DFT) fails in predicting the CDW and
superconducting orders.

In this work we demonstrate that the complete 
TiSe$_2$ pressure phase-diagram
is reproduced by first principles calculations. We 
show that the disappearing of the CDW 
at finite applied pressure (P) is due to the local chemistry around
a Ti atom while the pinning of the CDW is determined by the electron-phonon
coupling. Excitonic or nesting effects
are not the relevant interactions to describe the TiSe$_2$ pressure
 phase-diagram.
\begin{figure*}
\includegraphics[width=6.0cm]{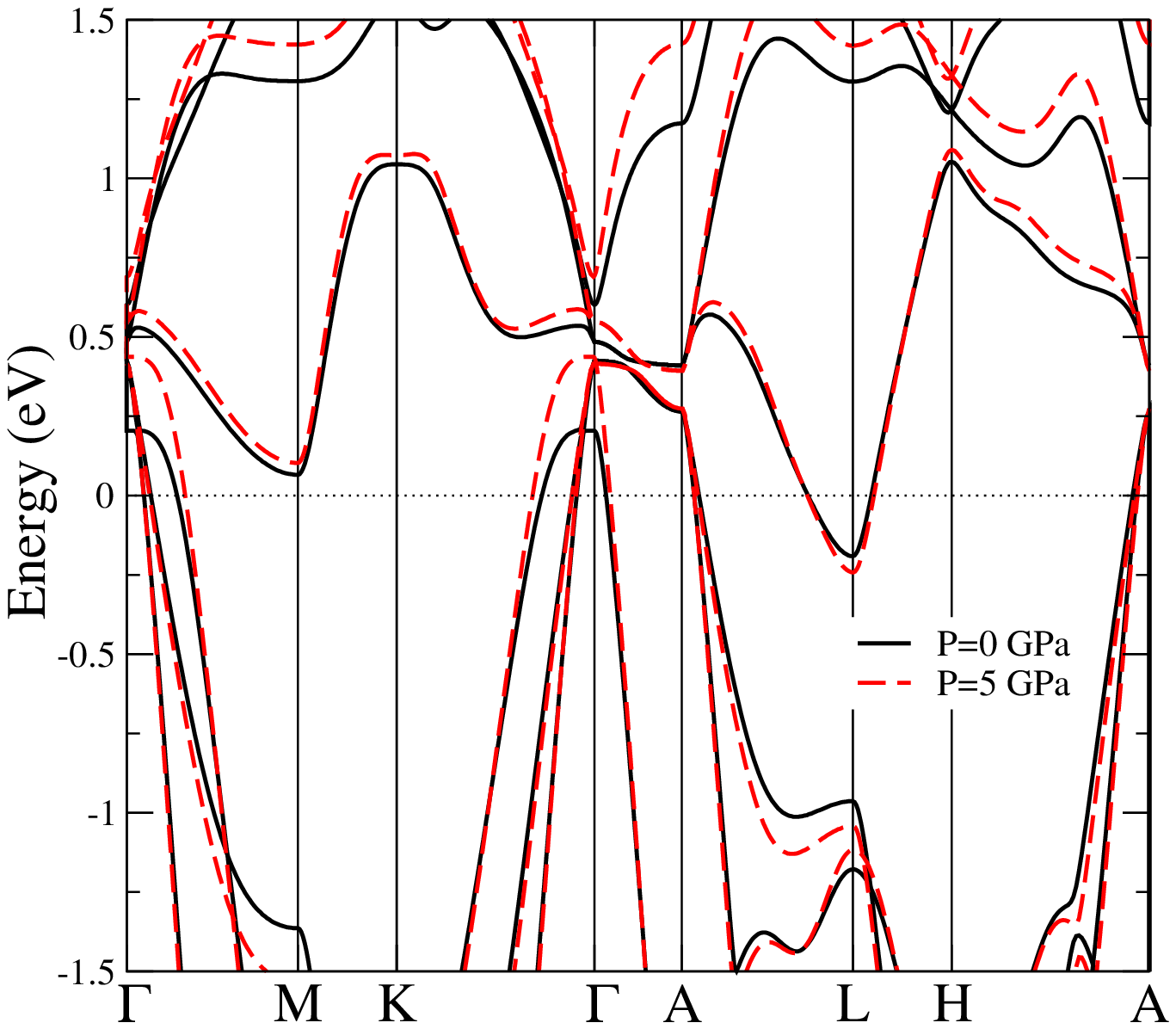}%
\includegraphics[width=6.0cm]{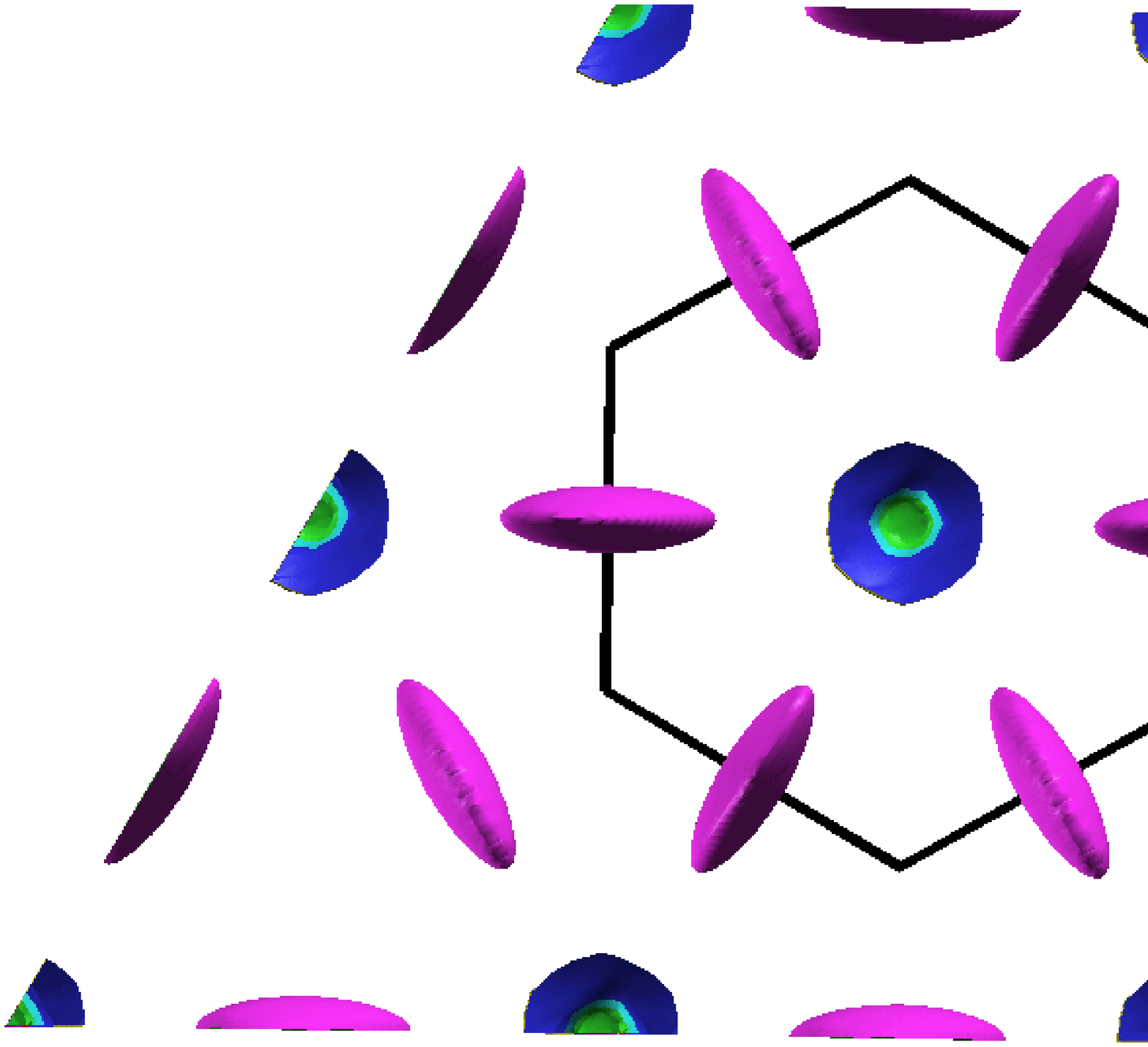}%
\includegraphics[width=6.0cm]{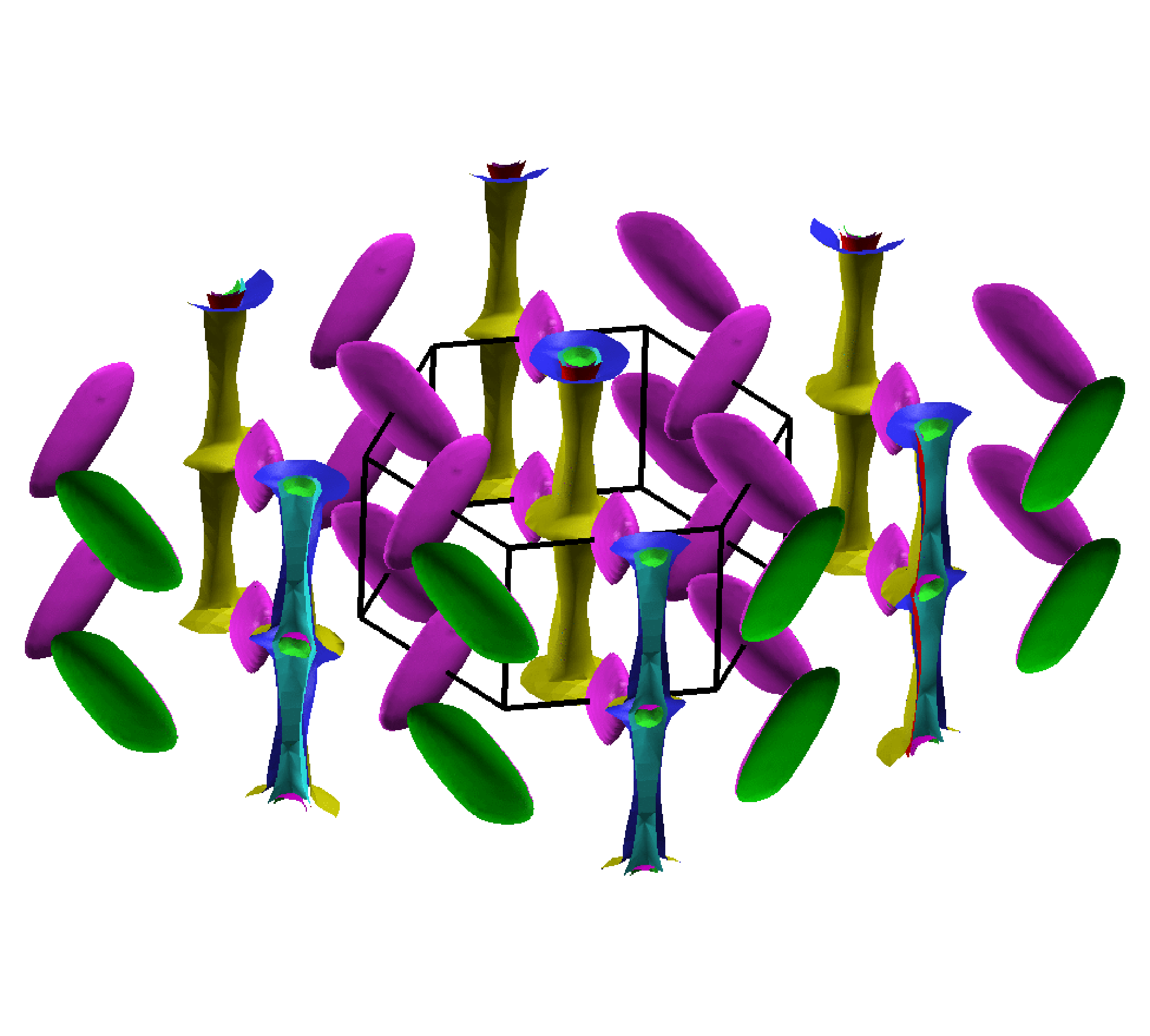}
\caption{(color online):(Left) TiSe$_2$ electronic structure at $P=0$ and $5$ GPa and (center and right) $P=0$ GPa Fermi surface.}%
\label{fig:el_struc}%
\end{figure*}

The DFT band structure \cite{Details} and Fermi surfaces of TiSe$_2$ at ambient pressure are
shown in Fig. \ref{fig:el_struc}. Our results are in agreement with previous works
\cite{Zunger78,Jishi2008}. 
Four bands cross the Fermi level. 
A narrow Ti-derived d-band 
is almost entirely unoccupied except around the L point where it
forms elongated and flat hole-pockets (in violet). 
A Se 4p derived band weakly hybridizes with Ti states and forms an
electron-pocket centered at $\Gamma$ having the form of a small
ellipsoid with larger axis parallel to
the k$_z=0$ plane. Finally the other two bands are formed by strongly
hybridized Se-p and Ti-d states.
Hydrostatic pressure weakly affects the band structure except for an increase
of the Se p band bandwidth.
\begin{figure}
\centerline{\includegraphics[width=0.9\columnwidth]{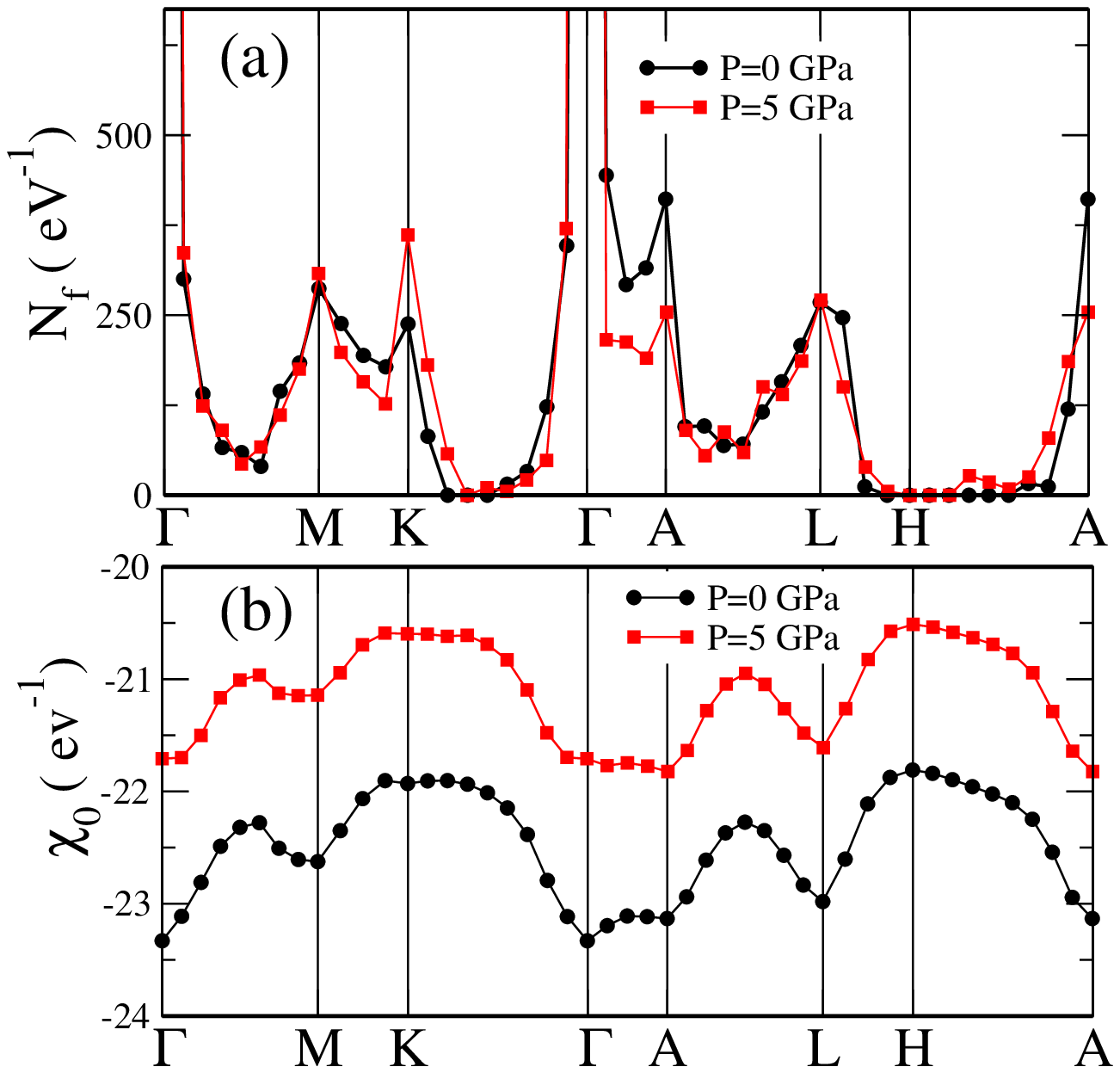}}
\centerline{\includegraphics[width=0.8\columnwidth]{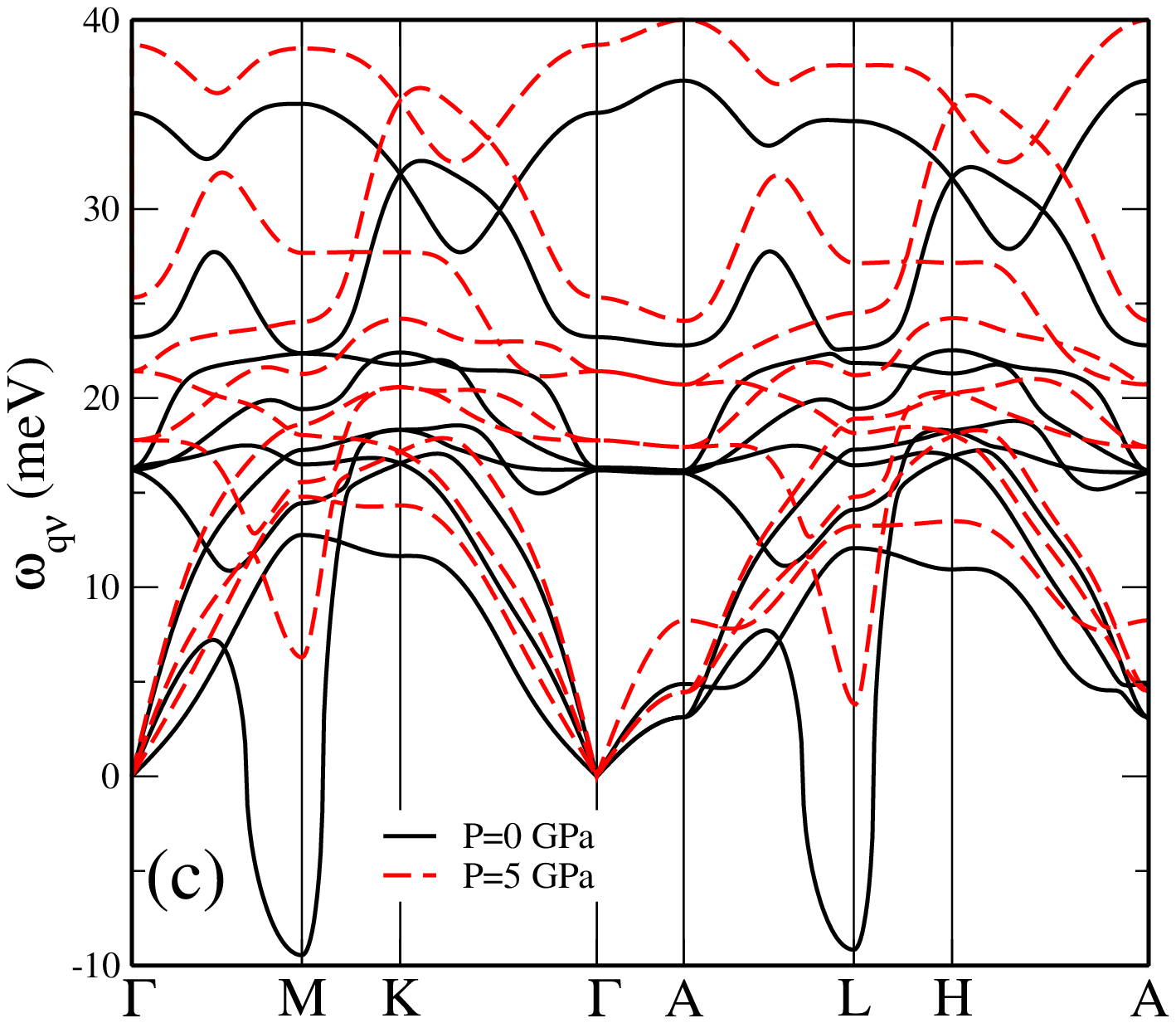}}
\centerline{\includegraphics[width=0.8\columnwidth]{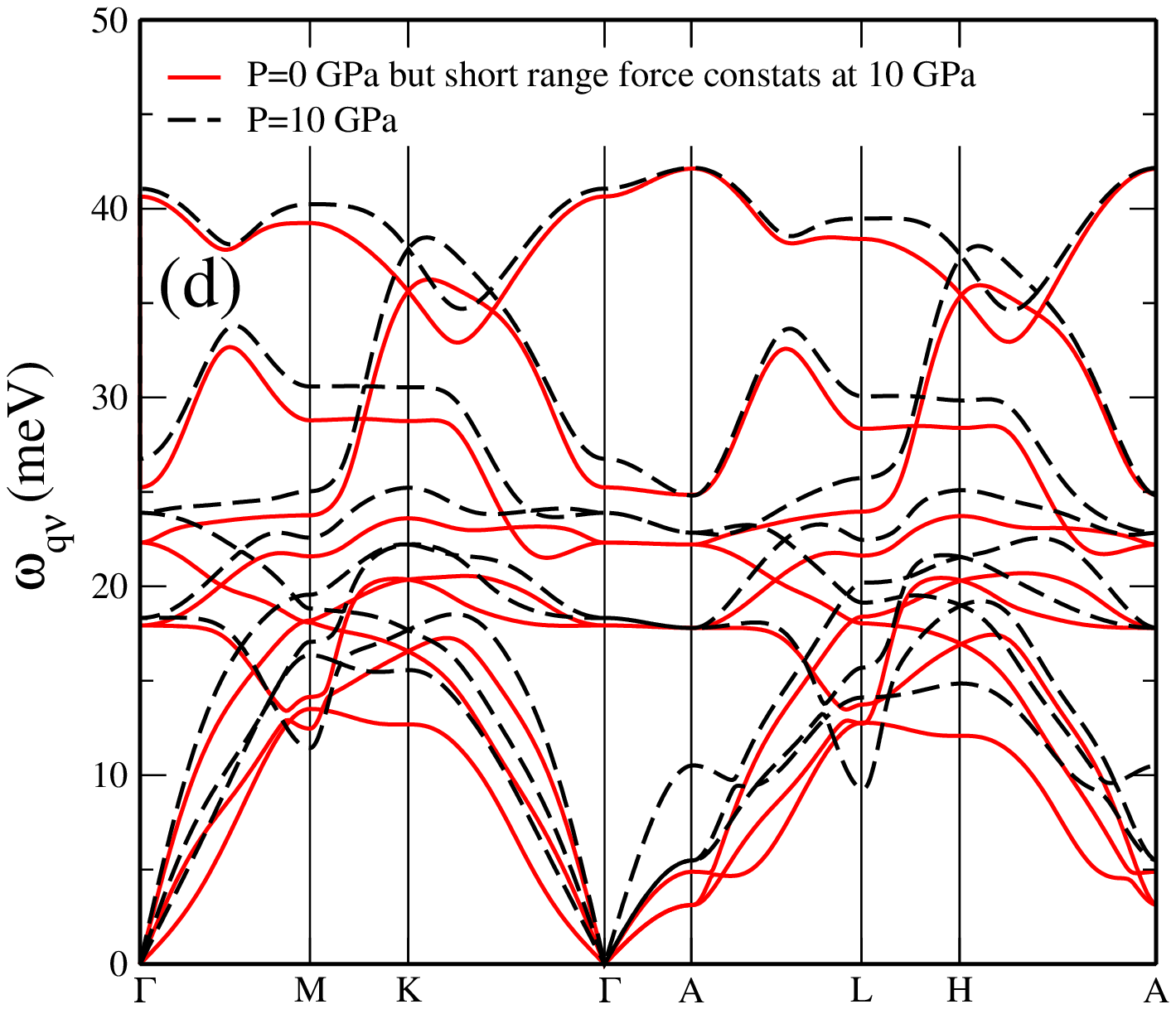}}
\caption{(color online) TiSe$_2$ (a) Nesting factor, (b) static susceptibility
 and (c) (d) phonon dispersion as a function  
hydrostatic pressure. In panel (d) the continuous line phonon dispersion
is obtained by replacing the nearest neighbours Ti-Se force constants
in the $P=0$ GPa force constant matrix with those at $P=10$ GPa. }%
\label{fig:ph_pressure}%
\end{figure}

The phonon dispersions \cite{Details} at pressure P=0,5 and 10 GPa are shown in Fig. 
\ref{fig:ph_pressure}. At zero pressure we find that the system is dynamically
unstable at the L and M points, consistent with a $2\times 2\times 1$ (M) or
a $2\times 2\times 2$ (L) real-space superstructure
in agreement with diffraction data \cite{MorosanNat2006}.
As pressure is increased, the CDW disappears at $P\approx 4$ GPa.
Transport data 
\cite{Li07} support
the occurrence of a superconducting state in the region between 2 and 4 GPa.
It is unclear if superconductivity coexist or not with CDW order.
Given that (i) anharmonic effects (neglected here) 
should be important close to the transition to the CDW state and 
should reduce the critical
pressure for the occurrence of CDW and (ii) the experimental phase diagram 
at finite pressure
is only based on transport measurements, our results are in excellent 
agreement with experimental finding. Thus DFT correctly
describes the occurrence of charge density way in TiSe$_2$ at zero and finite
pressure.

To address the microscopic mechanism responsible for the occurrence of CDW 
and its suppression under pressure we calculate the static bare susceptibility 
$\chi_0({\bf q})=\frac{2}{N_k} \sum_{{\bf k}nm}
\frac{f_{{\bf k}+{\bf q}n}-f_{{\bf k}m}}{\epsilon_{{\bf k}+{\bf q}n}-\epsilon_{{\bf k}m}}$
and the nesting factor 
$N_f({\bf q})=\frac{2}{N_k} \sum_{{\bf k}nm} \delta(\epsilon_{{\bf k}+{\bf q}n})
\delta(\epsilon_{{\bf k}m})$, where $\epsilon_{{\bf k}m}$ are 
the DFT-bands measured from the Fermi level.
 As $\chi_0({\bf q})$ coincides with the electron-phonon contribution to
the static phonon self-energy assuming constant matrix elements,
it represents the electron-phonon correction to the
phonon frequencies squared. A minimum in $\chi_0({\bf q})$ means a softening in
$\omega_{{\bf q\nu}}^2$. As shown in
Fig. \ref{fig:ph_pressure},  $\chi_0({\bf q})$ has a minimum both at M and L 
and $N_f({\bf q})$
 peaks at M and L. However both quantities are essentially
unaffected by pressure between 0 and 5 GPa, 
nesting cannot be responsible for
the disappearing of the CDW at 4 GPa.  

The mechanism responsible for the disappearance of CDW
under pressure can be understood by considering the 
force constants $C_{IJ}(P)$ at a given pressure $P$, where
$I,J$ label the atoms in the 
$4\times 4\times 2$ real-space supercell. 
A new set of short-range force constants 
$C_{IJ}^{SR}(P)$ is obtained
from $C_{IJ}(P)$ by setting to zero the force constants at distances larger than
the Ti-Se bond length. Then we define:
\begin{eqnarray}
{\tilde C}_{IJ}(P)&=&C_{IJ}(0)-C_{IJ}^{SR}(0)+C_{IJ}^{SR}(P).
\end{eqnarray}
The force-constants ${\tilde C}_{IJ}(P)$ and $C_{IJ}(0)$ only
differ by short-range terms as they both include the same long-range behavior
calculated at $P=0$.
From ${\tilde C}_{IJ}(P)$ we obtain new phonon frequencies at any 
phonon-momentum by Fourier interpolation. 
In Fig. 10 the phonon frequencies obtained with the
aforementioned procedure are compared with standard linear-response calculations.
As it can be seen the two are very similar and, most important, the CDW disappears
simply by removing the short-range force-constants at P=0 and replacing them
with that at $P=10$ GPa. The disappearing of the CDW in TiSe$_2$ is not driven by
any electronic mechanism or any long-range interaction but it is due to the stiffening of
the nearest-neighbors Ti-Se force-constants under applied pressure. 
The occurrence of a soft-phonon mode
is thus inherent to the local environment around the transition metal 
atom in the TiSe$_2$ layer. 

We then consider the superconducting properties of TiSe$_2$ 
at pressures $P\ge 5$ GPa. We calculate the
electron-phonon coupling $\lambda_{{\bf q}\nu}$ for a phonon mode
$\nu$ with momentum ${\bf q}$ and phonon frequency $\omega_{\bf q \nu}$, namely:
 
\begin{equation}\label{eq:elph}
\lambda_{{\bf q}\nu} = \frac{4}{\omega_{{\bf q}\nu}N(0) N_{k}} \sum_{{\bf k},n,m} 
|g_{{\bf k}n,{\bf k+q}m}^{\nu}|^2 \delta(\epsilon_{{\bf k}n}) \delta(\epsilon_{{\bf k+q}m}).
\end{equation}
The matrix element in Eq. \ref{eq:elph} is
$g_{{\bf k}n,{\bf k+q}m}^{\nu}= \langle {\bf k}n|\delta V/\delta u_{{\bf q}\nu} |{\bf k+q} m\rangle /\sqrt{2 \omega_{{\bf q}\nu}}$,
where
 $V$ is the Kohn-Sham potential and
$u_{{\bf q}\nu}$ is the amplitude of the phonon displacement
and $N(0)$ is the density of states at the Fermi level.
The critical temperatures obtained with McMillan formula and
the parameters in table  \ref{tab:elph}
are in qualitative agreement
with experiments although in experiments superconductivity occurs in a narrower 
range of pressures.
This discrepancy is probably related to the use of a 
McMillan type of approach to deduce the 
critical temperature and to the neglect of anharmonic effects.
\begin{table}
\begin{ruledtabular}
\begin{tabular}{c|c|c|c}
{\rm P(GPa)} & $\lambda$ &  $\omega_{\rm log}$(meV) & T$_c$(K) \\ \hline
5.0 & 1.57  &    7.0  &    0.84    \\
7.0 & 0.84  &   11.0  &    0.58    \\
10.0 & 0.62  &  15.1  &    0.38    \\
\end{tabular}
\end{ruledtabular}
\caption{Calculated superconducting properties of TiSe$_2$ using McMillan formula
$(\mu^{*}=0.1)$.  }
\label{tab:elph}
\end{table}
\begin{figure}
\includegraphics[width=0.8\columnwidth]{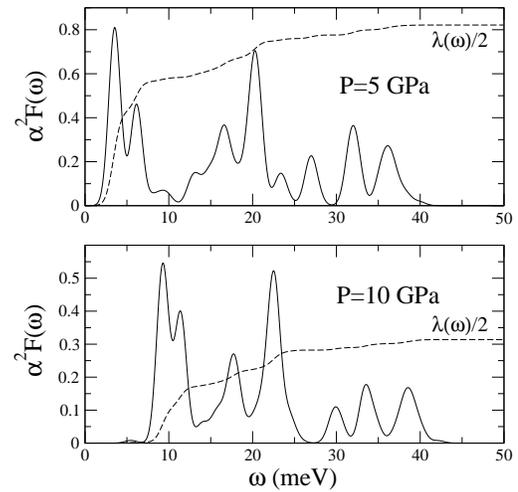}
\caption{ Eliashberg function and integrated Eliashberg function $\lambda(\omega)$
at 0 and 5 GPa }%
\label{fig:Elias}%
\end{figure}

We calculate the Eliashberg function
\begin{equation}
\alpha^2F(\omega)=\frac{1}{2 N_q}\sum_{{\bf q}\nu} \lambda_{{\bf q}\nu} \omega_{{\bf q}\nu} \delta(\omega-\omega_{{\bf q}\nu} )\label{eq:Elias}
\end{equation}
and the integral $\lambda(\omega)=2 \int_{0}^{\omega} d\omega^{\prime} 
\alpha^2F(\omega^{\prime})/\omega^{\prime}$. 
At $P=5$ GPa the Eliashberg function is dominated by the soft mode \cite{First_moment} at low energy that 
contributes to most of the coupling. As the pressure is raised, the low energy mode
hardens (see Fig. \ref{fig:ph_pressure}) and 
the average electron-phonon coupling is reduced, as shown
in Fig. \ref{fig:Elias}. 

In the present work we have shown that the finite pressure phase diagram 
of TiSe$_2$ can be entirely
explained by the local chemistry around a Ti atom and by the electron-phonon
interaction. In more details, the disappearing of the CDW at finite pressure
is determined by the stiffening of the 
Ti-Se nearest-neighbours short-range force-constants while the static bare electronic 
susceptibility determines the 
ordering vector at which the CDW locks-in. 
The electron-phonon interaction alone does not induce
a CDW, as it is clear from the 5 GPa calculation.

In previous works on other transition metal dichalcogenides
it was shown that nesting is indeed irrelevant in determining
CDW order \cite{Johannes06,Johannes08,Calandra2009} and that DFT
is able to reproduce the occurrence of CDW in 2H-NbSe$_2$ 
\cite{Calandra2009} and 1T-TaS$_2$ \cite{Liu2010}. In all these works the origin
of the CDW was attributed to the electron-phonon interaction. In our work
we show that, although electronic susceptibility is relevant in determining
the ordering vector, the local chemistry is the real responsible for
the occurrence of CDW. More exotic mechanisms such as an Overhauser transition
\cite{Li07} or excitonic phases \cite{Cercellier07} are unlikely.

Our work reproduces also the main features of the superconducting state of 
TiSe$_2$ as it explains the dome-like shape of Tc versus pressure
found in experiments \cite{KusmartsevaPRL2009}.
Superconductivity is driven by the softening
of the phonon mode related to the charge density wave. As the mode is very soft,
substantial anharmonic effects are to be expected. The inclusion of these
effects would harden the phonon frequency of the soft mode and shift the dome 
region in which superconductivity takes place to lower pressure in even better
agreement with experimental data. 

Several authors  \cite{MorosanNat2006,KusmartsevaPRL2009} pointed out
the strong similarity of the phase diagram of TiSe$_2$ with cuprates and
suggested a non-conventional pairing mechanism for TiSe$_2$.
Our work shows that this similarity is only apparent 
as here superconductivity is phonon-mediated.
Furthermore our work demonstrates that exotic phase diagrams including dome-like
superconducting regions, as those occurring in Cuprates, Pnictides or Cobaltates
can also arise from an electron-phonon mechanism. Thus, the apparent similarity of
these phase diagrams to the cuprate case cannot be considered a fingerprint of 
electronic correlation. 

Calculations were performed at the IDRIS supercomputing center (project
081202).


\begin{thebibliography}{99}
\bibitem{Details} Density functional theory (DFT) calculations are performed using the
Quantum-Espresso code \cite{QE} 
in the local density approximation (LDA)
\cite{PZ}. Norm-conserving\cite{Troullier} and Ultrasoft \cite{Vanderbilt}
pseudopotentials are used for
 Se and Ti, respectively. 
The phonon dispersion and the electron-phonon \cite{Details} coupling are calculated
using density functional perturbation theory in the linear 
response\cite{BaroniRMP}. A
$N_{k}=24\times 24\times 12$  grid and an Hermite-Gaussian smearing
of $0.01$ Ryd are used for the
electronic integration. A 
$N_{k}=40\times 40\times 24$ grid is used for the electron-phonon
coupling. The dynamical matrices are calculated on a 
$N_{q}=4\times 4\times 2$ phonon-momentum grid and are Fourier interpolated
throughout the full Brillouin zone.
The electronic bare susceptibility
is calculated using maximally localized Wannier functions\cite{Marzari,Mostofi}
and a $N_k=75\times 75\times 25$ electron-momentum grid.
We perform structural optimization at finite pressure rescaling the
theoretical LDA \cite{PZ} 
volume with respect to the $P=0$ GPa one.
\bibitem{First_moment}
The first momentum of the Eliashberg function, namely
$M_1=\int_{0}^{\infty} \alpha^2F(\omega) \omega d\omega$, is independent
of the phonon frequency. As $M_1(5\, {\rm GPa})=1.35\times M_1(10\, {\rm GPa})$ 
while $\lambda( 5\, {\rm GPa})=2.54\times \lambda(10 \,{\rm GPa})$,
the decrease of $M_1$ with pressure is much smaller than that of $\lambda$.
The main effect is then the stiffening
of the Ti-Se force constants resulting in an hardening of the lowest energy phonon mode.
\end{thebibliography}
\end{document}